\documentclass[twocolumn]{aastex631}

\usepackage{fancyvrb}

\usepackage{svg}
\usepackage{tikz}

\newcommand{\sabinasays}[1]{{\color{purple}#1}}

\usepackage{amsmath} 
\usepackage{multirow} 
\usetikzlibrary{shapes.geometric, arrows}
\undef\listofchanges
\usepackage[commandnameprefix=always,commentmarkup=uwave]{changes}
\definechangesauthor[name={Will Farr}, color=cyan]{wf}

\shorttitle{Stellar rotation in NGC~6819}
\shortauthors{Sagynbayeva, Colman, \& Farr}
\graphicspath{{./}{figures/}}
\begin{document}

\title{Rotation Periods for Stars in Open Cluster NGC~6819 From \textit{Kepler} IRIS Light Curves}

%% you desire. In v6.31 you can include a footnote in the title.

%% A significant change from earlier AASTEX versions is in the structure for 
%% calling author and affiliations. The change was necessary to implement 
%% auto-indexing of affiliations which prior was a manual process that could 
%% easily be tedious in large author manuscripts.
%%
%% The \author command is the same as before except it now takes an optional
%% argument which is the 16 digit ORCID. The syntax is:
%% \author[xxxx-xxxx-xxxx-xxxx]{Author Name}
%%

\correspondingauthor{Sabina Sagynbayeva}
\email{sabina.sagynbayeva@mcgill.ca}

\author[0000-0002-6650-3829]{Sabina Sagynbayeva}
\altaffiliation{CITA National Fellow}
\affiliation{Department of Physics and Astronomy, Stony Brook University, Stony Brook, NY 11794, USA}
\affiliation{McGill University, Trottier Space Institute, 3550 rue University, Montreal, QC, H3A 2A7, CAN}

\author[0000-0001-8196-516X]{Isabel L. Colman}
\affiliation{Liberal Studies, New York University, 726 Broadway, New York, NY 10003, USA}
\affiliation{Department of Astrophysics, American Museum of Natural History, 200 Central Park West, New York, NY 10024, USA}

\author[0000-0003-1540-8562]{Will M. Farr}
\affiliation{Department of Physics and Astronomy, Stony Brook University, Stony Brook, NY 11794, USA}
\affiliation{Center for Computational Astrophysics, Flatiron Institute, 162 Fifth Avenue, New York, NY 10010, USA}



\begin{abstract}
We present an updated catalog of stellar rotation periods for the 2.5 Gyr open cluster NGC 6819 using the \textit{Kepler} IRIS light curves from superstamp data. We present two modeling techniques: one for measuring rotation periods from light curves in crowded fields, and one for fitting a gyrochronology relation. Our analysis uses Gaussian Process modeling to extract robust rotation signals from image subtraction light curves, allowing us unprecedented data access and measurement precision in the crowded cluster field. After applying stringent quality and contamination cuts, we identify 152 reliable rotation periods. Compared to previous work, which relied on only $\sim$30 stars, our catalog extends the gyrochronological sequence of NGC 6819 with an order of magnitude more measurements and improved precision. The expanded dataset reveals both the expected temperature-dependent spin-down trend and substantial scatter at fixed effective temperature, including a tail in the period distribution revealing some abundance of fast rotators. These results strengthen this cluster's role as a benchmark for stellar spin evolution, while also highlighting the limitations of traditional gyrochronology at older ages. The final catalog and the model implementations are all available on Zenodo\footnote{see \url{https://doi.org/10.5281/zenodo.23431597}}. 

\end{abstract}



\section{Introduction} \label{sec:intro}

Accurate stellar age determination remains one of the most significant challenges in stellar astrophysics. Gyrochronology, the connection between stellar surface rotation and age \citep{Barnes2003, Barnes2007, Mamajek2008}, provides a viable method for age estimation along the main sequence. The fundamental assumption underlying gyrochronology --- that stellar rotation periods can serve as reliable age indicators --- is increasingly called into question by observational evidence that stellar spin evolution is far more complex than previously thought. Traditional gyrochronology models assume that after an initial convergence phase around 500-700 Myr, stars of similar mass and metallicity will follow predictable spin-down trajectories governed by magnetic braking, making the rotation period a monotonic function of age for single main-sequence stars \citep{Barnes2003, Mamajek2008}.

However, multiple lines of observational evidence now challenge this paradigm. \cite{Angus2020} used velocity dispersions as kinematic age proxies to demonstrate that this framework breaks down for stars older than $\sim1$ Gyr, showing that stars with identical rotation periods and colors can have vastly different ages. This suggests that some physical processes, like variable core-envelope angular momentum transport, or changes in magnetic field topology, can fundamentally alter rotational evolution in ways that current gyrochronology calibrations cannot capture. Supporting this conclusion, \cite{vanSaders2016} found evidence for weakened magnetic braking in stars with Rossby numbers greater than $\sim2$, approximately corresponding to the Sun, which would cause a dramatic slowdown in rotational evolution for older, slowly rotating stars.

Understanding the pre-main-sequence context is also crucial for interpreting the main-sequence challenges. \cite{Matt2010} demonstrated that magnetic star-disk coupling plays a critical role in setting initial main-sequence rotation rates, but the effectiveness of this ``disk locking" mechanism depends sensitively on uncertain parameters, including magnetic field strength and the degree of magnetic coupling to the disk. When \cite{Matt2010} include field-line opening due to differential rotation between star and disk (rather than assuming a fully closed field), they find that magnetic torques alone cannot explain young stars observed with surface rotation periods of 1--10 days. This suggests that additional angular momentum loss mechanisms --- particularly stellar winds --- may be necessary even during the disk accretion phase, which would further complicate the initial conditions for main-sequence gyrochronology. 
Further, \cite{Matt2015} demonstrated that the mass-dependence of angular momentum evolution in Sun-like stars, with lower-mass stars spinning down faster than solar-mass stars once in the magnetically unsaturated regime, creating the observed gyrochronology sequence, can only be reproduced by incorporating empirically-derived torque scalings that depend on both stellar mass and the Rossby number. This highlights how sensitively rotational evolution depends on magnetic field properties and wind mass-loss rates. Open clusters, where all members are assumed to be co-eval, provide a laboratory to search for signatures of these underlying physics, and build on the past two decades of work aimed at characterizing the angular momentum evolution of Sun-like stars.

Thanks to high-quality data from NASA's Kepler mission \citep{Borucki2010}, NGC 6819 serves as one of the key gyrochronology calibrator clusters \citep[e.g.,][]{Bouma2023, Curtis2020}. With an age of 2.5 Gyr \citep{Hole2009, Anthony-Twarog2014} and solar metallicity \citep[Z = $0.09 \pm 0.03$;][]{Lee-Brown2015}, this cluster has so far provided crucial constraints for gyrochronological relations in the intermediate age regime. Currently, the stellar rotation periods derived by \cite{Meibom2015} serve as the standard reference for this cluster.
While the establishment of the NGC 6819 gyrochronology sequence represents an important milestone, the sample comprises only approximately 30 stars, and now, with the availability of more light curves for this cluster \citep{Colman2022}, we have the ability to conduct a more complete investigation and strengthen gyrochronological calibrations relying on results from this cluster. We seek to independently verify or refine these detections of surface rotation, as well as to expand the sample size through additional detections.

During the \textit{Kepler} mission, the extreme stellar crowding characteristic of cluster environments limited standard pipeline processing to postage stamp cutouts for only a small number of bright, isolated stars in NGC~6819, many of which were not confirmed cluster members. To address this limitation, NGC 6819, along with NGC 6791, was targeted with ``superstamps" --- 200 square pixel images centered on the cluster core that were downloaded at \textit{Kepler}'s long cadence \citep{Jenkins2010}.

Extracting stellar rotation periods from \textit{Kepler} photometry presents unique data analysis challenges. Standard \textit{Kepler} pipeline light curves cannot be used directly for rotation studies, as the data processing was optimized for exoplanet detection and likely removes or suppresses the photometric variability associated with stellar rotation \citep{Smith2012, Reinhold2013}. This limitation does not affect superstamp observations, since custom photometry has to be performed regardless. This work uses the \textit{Kepler} IRIS light curves \citep{Colman2022}, which represent a comprehensive photometric analysis of the NGC 6819 superstamp data. Several previous studies have successfully leveraged \textit{Kepler} superstamp observations for cluster studies \citep[e.g.,][]{Covelo-Paz2023, Kuehn2015, Sanjayan2022}. The IRIS light curve extraction method is described in the data release paper, and our data reduction methodology is detailed in Section \ref{sec:data}. 

\textit{Kepler}'s 90-day quarter structure provides a sufficient temporal baseline for rotation period detection. Experience with TESS observations demonstrates that data downlink gaps present a more significant obstacle than sector duration limitations \citep{Colman2022, Holcomb2022}, indicating that even a single \textit{Kepler} quarter can theoretically accommodate rotation periods up to 45 days without requiring complete phase coverage. While longer periods could be accessed through careful stitching of multiple quarters, this approach is unnecessary for the present study.
Stellar rotation theory predicts that main-sequence stars in a 2.5 Gyr cluster should exhibit rotation periods no longer than approximately 30 days \citep{Meibom2011}, well within \textit{Kepler}'s detection capabilities. However, the 4-year \textit{Kepler} baseline is sensitive to magnetic activity cycles \citep{Reinhold2017}, which can manifest as periodic signatures in highly active stars as starspot distributions migrate in latitude \citep[e.g.,][]{Reinhold2015, Morris2017, Sagynbayeva2025}. Our approach in this work is designed specifically to detect rotational modulation (see Section~\ref{sec:methods}), and our sample is unlikely to be polluted by detections of activity cycles, as these tend to be on the scale of months to years long for FGK dwarfs \citep[e.g.,][]{erika2007}.

With this expanded photometric dataset, we revisit NGC 6819 to provide an updated catalog of stellar rotation periods, increasing the sample size from the previous 30 detections \citep{Meibom2015} to rotation periods. A comprehensive analysis of the rotational properties and their implications for gyrochronology is presented in Section \ref{sec:results}. We present these measurements to the community as a refined 2.5 Gyr gyrochronological sequence, providing improved statistical constraints for age-rotation relationships. Our methodology demonstrates robust performance for rotation period detection in \textit{Kepler} photometry and shows promise for application to other high-precision photometric surveys, including TESS \citep{Ricker2015} and future missions such as PLATO \citep{Rauer2014}. This work also highlights the scientific potential of \textit{Kepler}'s superstamp observations for stellar cluster studies. 

Our paper is organized as follows: we present the data in Section \ref{sec:data}; we show the details of our models for co-trending, period determination, and gyrochrone fitting in Section \ref{sec:methods}; we provide a detailed overview of the results and a discussion in Section \ref{sec:results}; and finally present our conclusions in Section \ref{sec:conclusions}.

\section{Data} \label{sec:data}
We used light curves from the \textit{Kepler} IRIS catalog \citep{Colman2022, irismast}, which provides data for 9,150 stars in and around the open clusters NGC~6791 and NGC~6819, publicly available as a high-level data product on the Mikulski Archive for Space Telescopes (MAST)\footnote{https://archive.stsci.edu/hlsp/iris}. These light curves were produced using a novel implementation of image subtraction photometry on \textit{Kepler} ``superstamp" data. The catalog includes 8,427 stars that were never targeted during the original Kepler mission, plus additional quarters of data for 382 previously targeted stars. The IRIS (Increased Resolution Image Subtraction) method follows the traditional image subtraction approach of treating all non-variable signals as background \citep{Alard1998}, which optimizes light curves for variable signal, with the additional step of upsampling \textit{Kepler} pixel data to enable more accurate image alignment before background subtraction. This enables the extraction of high-quality photometry from crowded stellar fields, and is also particularly sensitive to faint targets.

\textit{Kepler}'s coverage was not continuous throughout its four-year mission due to the failure of Module 3 during Quarter 4, which reduced the number of operational CCD channels from 84 to 80 \citep{kepler2016}. This region included NGC 6819, meaning that all of our targets are missing Quarters 6, 10, and 14. Additional targets received only partial coverage due to detector positional variation between rotations, which affected a small number of targets towards the edge of the superstamps.

For cluster membership determination, we utilized the astrometric membership probabilities combined from the WIYN Open Cluster Study catalog of NGC 6819 \citep{Platais2013} and the \textit{Gaia DR2} catalog reported in \cite{Gaudin2018}, which provides proper motion-based membership analysis for 15,750 stars in the cluster field with accuracies ranging from $\sim$0.2 mas yr$^{-1}$ in the cluster center to 1.1 mas yr$^{-1}$ in the outer regions. This membership list, combined with the IRIS catalog, gave us 2,288 stars in the cluster, to which we applied our analysis.

% %
% \begin{figure}[hbt!]
%     \center
%     \includegraphics[width=0.46\textwidth]{example-lc-notdetrend.png}
%         \caption{An example of some IRIS light curves before co-trending.}
%         \label{fig:lcs-iris}
% \end{figure}
% %

\section{Methods} \label{sec:methods}
\subsection{Co-trending of \textit{Kepler} Light Curves}

The \textit{Kepler} IRIS light curves are provided as raw data for each quarter, or one stitched and co-trended time series; however, the initial co-trending was optimized for asteroseismology at the expense of longer simple-periodic signals. As part of the \textit{Kepler} mission, official data releases also provided co-trending basis vectors (CBVs) used to perform data reduction \citep{kepler2016}. The use of CBVs is dependent on the location of each target on the detector; since the majority of our targets were not included in official \textit{Kepler} data releases, and in the interests of uniformity, we followed a similar procedure in order to co-trend the raw IRIS data. We sought to identify a common set of basis vectors that provides an optimal linear fit to our ensemble of light curves. This approach allows us to construct a design matrix $M$ that captures the dominant systematic trends present across the dataset (see some examples of CBVs in the Appendix \ref{sec:cbvs}).

For a given design matrix $M$, we fit each light curve $y_i$ (where $i = 1, \ldots, N$) using coefficients $x_i$ by maximizing the likelihood:

\begin{equation}
\log \mathcal{L}_i = -\frac{1}{2} \left( y_i - M x_i \right)^T \Sigma_i^{-1} \left( y_i - M x_i \right) + \mathrm{const},
\end{equation}
where $\Sigma_i$ is the covariance matrix of the noise in lightcurve $y_i$.

The optimal likelihood is achieved when we project out the component of $y_i$ that lies within the column space of $M$. This optimal likelihood is given by:

\begin{equation}
\log \hat{\mathcal{L}}_i = -\frac{1}{2} \left(y^{\perp}_i \right)^T \Sigma_i^{-1} \left( y^{\perp}_i \right) + \mathrm{const}
\end{equation}

where $y^{\perp}_i$ represents the component of $y_i$ orthogonal to the columns of $M$. Through optimization of the likelihood with respect to $x_i$, we obtain:

\begin{equation}
y^{\perp}_i = \left( I - M \left( M^T \Sigma_i^{-1} M \right)^{-1} M^T \Sigma_i^{-1} \right) y_i
\end{equation}

The general problem of finding an optimal $M$ would require sampling within this
model; but here we make the reasonable assumption that each lightcurve is
subject to white (i.e. independent, identically-distributed) noise, whose
amplitude we estimate empirically in the following way.  First, we difference
the lightcurve to eliminate any long-timescale systematic or stellar trends;
then we compute the variance of the differenced lightcurve, and estimate the
variance of the lightcurve white noise by dividing by two.  Thus $\Sigma_i =
\sigma^2_i I$, where $\sigma^2_i$ is the estimated white noise variance of
lightcurve $i$, and $I$ is the identity matrix.

In this simplified scenario, the optimal likelihood for a given design matrix
$M$ is:
\begin{multline}
\log \hat{\mathcal{L}}_i = -\frac{1}{2 \sigma_i^2} \left( y^{\perp}_i \right)^T y^{\perp}_i + \mathrm{const} \\ = -\frac{1}{2} \left( \tilde{y}^{\perp}_i \right)^T \left( \tilde{y}^{\perp}_i \right) + \mathrm{const}
\end{multline}
where $\tilde{y}_i = y_i / \sigma_i$ represents the standardized light curve with unit noise variance. 
% Figure \ref{fig:lcs-iris} shows some standardized light curves before co-trending. 
The total likelihood across all light curves is obtained by summation:
\begin{equation}
\log \hat{\mathcal{L}} = \sum_i \log \hat{\mathcal{L}}_i = -\frac{1}{2} \sum_i \left( \tilde{y}^{\perp}_i \right)^T \left( \tilde{y}^{\perp}_i \right) + \mathrm{const}
\end{equation}

To determine the optimal design matrix $M$ that maximizes the likelihood, we employ singular value decomposition (SVD). The objective is to identify the subspace of whitened light curves that contains the maximum squared length for a given number of components in $M$. The complementary subspace will minimize the sum-of-squares over any $(N-k)$-dimensional subspace.

We are thus led to compute the SVD of the whitened light curves and select the
first $k$ singular vectors to form the co-trending basis that becomes the
columns of $M$.  This is similar to the procedure used in the \textit{Kepler}
pipeline to compute CBVs \citep{kepler2016}, but that procedure did not whiten
before computing singular vectors, and so likely over-weights the noisiest
lightcurves in the estimation of CBVs; our procedure is optimal under the
assumption that (1) the noise is independent and identically distributed and
well-characterized by the estimated variance and (2) instrument systematics
enter each lightcurve as a linear combination of common ``systematic modes.''
Figure \ref{fig:pQCBV2} shows the CBVs that result from our procedure for all
quarters of observations.

To ensure proper treatment of the mean flux, we zero-mean all whitened light curves before applying SVD. This procedure guarantees that: (1) the constant vector has zero singular value, and (2) the first $k$ singular vectors are orthogonal to the constant vector. We then augment the $k$ singular vectors with the constant vector to construct a $(k+1)$-dimensional design matrix that simultaneously fits the mean flux and removes the estimated linear trends.

%
% \begin{figure}[hbt!]
%     \center
%     \includegraphics[width=0.46\textwidth]{CBVs.png}
%         \caption{A subset of 9 different CBVs.  \sabinasays{This figure and caption should be changed to show (1) the actual number of CBVs we used for detrending in this quarter (which should be identified).  It should say that the CBVs show obvious long-timescale systematics (ie they don't look "noisy"), and we should plot all the CBVs in the appendix for all quarters (and say how many we used in each quarter).}}
%         \label{fig:cbvs}
% \end{figure}
%

\begin{figure*}[hbt!]
    \center
    \includegraphics[width=0.9\textwidth]{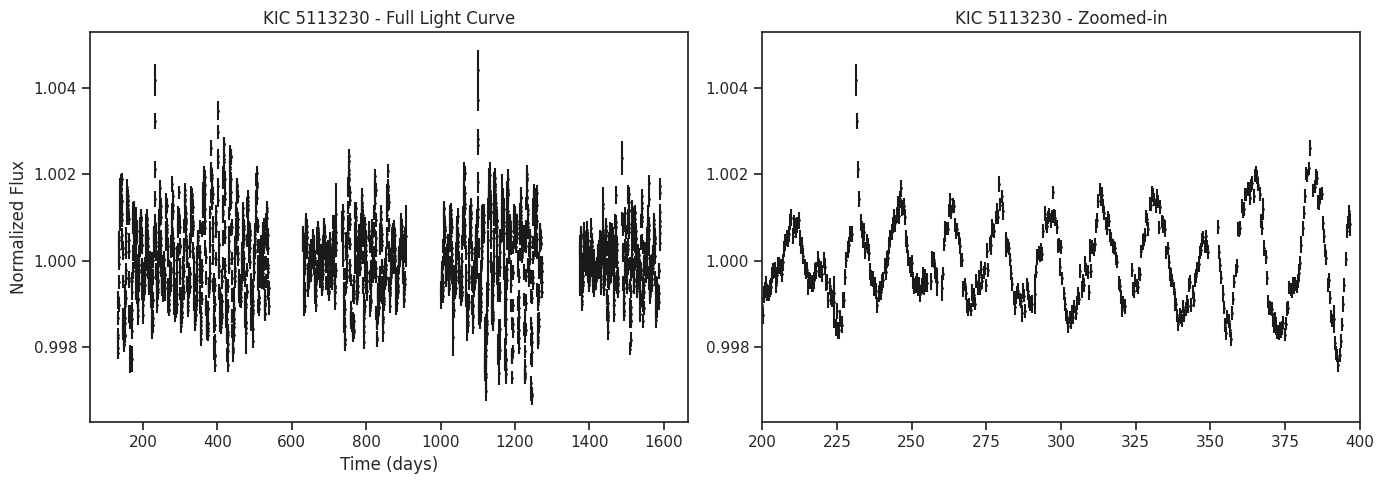}
        \caption{An example of a light curve after our co-trending. We also binned the light curves for computational efficiency. The left panel shows the full light curve, and the right panel shows the zoomed-in light curve between 200 and 400 days. We show each quarter's flux relative to \emph{that quarter's} mean flux. Therefore, the apparent changes in the fluctuations quarter-to-quarter are not necessarily real, as the mean flux will vary from quarter to quarter based on the instrument sensitivity.  The model described in Section \ref{sec:gp} accounts for this, assigning the (common-to-all-quarters) GP to the fractional flux variations, with a per-quarter mean flux that is fit to the data. We show more of the light curves with the period posterior distributions in the Appendix \ref{sec:apxlc}}
        \label{fig:detrended}
\end{figure*}

It remains to choose the number of CBVs to incorporate in the co-trending; the
likelihood above provides no direct information on the optimal number, since the
optimal likelihood is achieved by including every basis vector spanning the
entire collection of lightcurves.  We chose to include additional CBVs until the
coefficient of the last-added CBV in the median star reached 5-sigma
significance above zero.  Thus the last-added CBV meaningfully improves the
likelihood of the median star.  This procedure leads to $\mathcal{O}(8)$ CBVs in
each quarter.  De-trended lightcurves lack the dramatic systematic trends, but some systematic effects remain;
these will be handled by our low-frequency DRW term in the rotation kernel
discussed in the next subsection. We bin the final co-trended light curves by the
bin-width of $0.4 \, \mathrm{d}$ for computational efficiency.  An example of an
analysis-ready light curve appears in Figure \ref{fig:detrended}.

\subsection{The Gaussian Process Model for Rotation Periods}
\label{sec:gp}
We implement a Gaussian Process (GP) model to characterize stellar rotation
signatures in multi-quarter photometric time series, first introduced in
\citet{Foreman-Mackey2017} as a ``rotation kernel.'' The model represents the
stellar variability as a sum of harmonic oscillations corresponding to the
fundamental rotation frequency and its first harmonic.  We add to it terms
approximating a damped-random-walk (``real term'') at low frequency to account
for remaining long-timescale systematics \citep{Foreman-Mackey2017}.

For each observation $i$ at time $t_i$ in quarter $q_i$, we model the observed flux $y_i$ as:
\begin{equation}
y_i = \mu_{q_i} \cdot (1 + \text{GP}(t_i)) + \epsilon_i
\end{equation}
where $\mu_{q_i}$ is the mean flux level for quarter $q_i$, $\text{GP}(t_i)$ represents the fractional variability described by a Gaussian Process, and $\epsilon_i \sim \mathcal{N}(0, \sigma_{err,i}^2)$ is the measurement noise.

To account for instrumental variations and systematic effects between observing quarters, we introduce quarter-dependent scaling factors. The mean flux for each quarter is modeled as:
\begin{equation}
\mu_q = \bar{\mu}_q \cdot s_q
\end{equation}
where $\bar{\mu}_q$ is the empirical mean flux in quarter $q$ and $s_q$ is a scaling factor. The log-scaling factors are assigned hierarchical priors:
\begin{equation}
\log s_q \sim \mathcal{N}\left(0, \left(\frac{\hat{\sigma}_q}{\sqrt{n_q}}\right)^2\right)
\end{equation}
where $\hat{\sigma}_q$ is the relative standard deviation of flux measurements in quarter $q$ and $n_q$ is the number of observations in that quarter.

The GP kernel consists of two components designed to capture different aspects of stellar variability. The primary component is a rotation term that captures quasi-periodic stellar variability, while a secondary component models longer-timescale stochastic variations that could otherwise interfere with period detection.

The rotation kernel employs a \texttt{RotationTerm} from \texttt{celerite} that combines two damped harmonic oscillators to represent the fundamental rotation frequency and its first harmonic:
\begin{multline}
k_{\text{rot}}(\tau) = \sigma^2 \left[ \frac{1}{2}(1+f) \exp\left(-\frac{|\tau|}{Q_0 P}\right) \cos\left(\frac{2\pi|\tau|}{P}\right) \right. \\
+ \left. \frac{1}{2}(1-f) \exp\left(-\frac{|\tau|}{Q_1 P}\right) \cos\left(\frac{4\pi|\tau|}{P}\right) \right]
\end{multline}
where:
\begin{itemize}
\item $\sigma^2$ is the variance amplitude
\item $P$ is the rotation period
\item $Q_0$ and $Q_1$ are quality factors controlling the damping of the fundamental and first harmonic
\item $f \in [0,1]$ controls the relative amplitude of the two harmonics
\item $\tau = |t_i - t_j|$ is the time lag
\end{itemize}

To prevent the rotation kernel from fitting spurious long-period variations that could mask the true stellar rotation signal, we include an additional stochastic component modeled as a Simple Harmonic Oscillator (SHO) term operating in the heavily damped regime ($Q = 0.1$). This component effectively acts as a \texttt{RealTerm} in the power spectral density, providing a mechanism to absorb red noise and long-timescale trends:

\begin{equation}
k_{\text{red}}(\tau) = \text{SHOTerm}(S_0, \omega_0, Q=0.1)
\end{equation}

The parameters of this red noise component are constrained based on the observation characteristics:
\begin{align}
\tau_{\text{longest}} &= \frac{1}{f_0/\sqrt{2}} \\
\log \tau_c &\sim \text{Uniform}(\log \tau_{\text{longest}}, \log T_{\text{obs}}) \\
\omega_0 &= \frac{2\pi/\tau_c}{2Q} \\
S_0 &= \frac{\sigma_{\text{red}}^2}{\omega_0 Q}
\end{align}
where $\tau_c$ is the characteristic timescale, $T_{\text{obs}}$ is the total observation duration, and $\sigma_{\text{red}}$ controls the amplitude of the red noise component. The timescale is constrained to be longer than the expected rotation period but shorter than the total observation baseline, ensuring that this component captures long-term variations without interfering with the periodic signal of interest.

The total kernel is the sum of these components:
\begin{equation}
k(\tau) = k_{\text{rot}}(\tau) + k_{\text{red}}(\tau)
\end{equation}

The kernel's power spectral density is computed at specified frequencies $\omega$:
\begin{equation}
S(\omega) = \int_{-\infty}^{\infty} k(\tau) e^{-i\omega\tau} d\tau
\end{equation}

\subsubsection{Priors}
The model parameters are assigned the following priors:
\begin{equation}
\log P \sim \mathcal{N}(-\log f_0, \sigma_f^2)
\end{equation}
where $f_0$ is the prior guess for the fundamental frequency and $\sigma_f$ is the fractional uncertainty.
\begin{equation}
\sigma \sim \text{HalfNormal}(0.1 \cdot \text{median}(\hat{\sigma}_q))
\end{equation}
\begin{equation}
\Delta Q_0, \Delta Q_1 \sim \text{LogNormal}(\log 5, 1)
\end{equation}
with $Q_0 = 0.5 + \Delta Q_0 + \Delta Q_1$ and $Q_1 = 0.5 + \Delta Q_1$.
\begin{equation}
f \sim \text{Uniform}(0, 1)
\end{equation}

For the red noise component:
\begin{equation}
\sigma_{\text{red}} \sim \text{HalfNormal}(0.1 \cdot \text{median}(\hat{\sigma}_q))
\end{equation}

The centered and scaled observations are defined as:
\begin{equation}
\tilde{y}_i = \frac{y_i}{\mu_{q_i}} - 1
\end{equation}

The GP likelihood is computed for these transformed observations:
\begin{equation}
\log p(\tilde{y} | \theta) = -\frac{1}{2} \tilde{y}^T K^{-1} \tilde{y} - \frac{1}{2} \log |K| - \frac{n}{2} \log(2\pi)
\end{equation}
where $K$ is the covariance matrix derived from the combined kernel.

A Jacobian correction accounts for the transformation from $y$ to $\tilde{y}$:
\begin{equation}
\log p(y | \theta) = \log p(\tilde{y} | \theta) - \sum_{i=1}^n \log \mu_{q_i}
\end{equation}

The model has difficulty ``transitioning'' between multiple modes in period space if the range of allowed rotation periods is too large; it tends to get stuck, for example, with the first harmonic assigned to the fundamental frequency, or with the fundamental assigned to a first harmonic.  Transitions away from these local maxima require large changes in the parameters, and therefore have low probability.  To address this, we run the model multiple times with different priors on the rotation period, always constraining the range of the fundamental to a factor of two in period space, ensuring that, within that prior range, there is no secondary mode in the likelihood that could trap the sampler.

For each star, we ran the model with five priors on frequency (inverse of period) corresponding to five ranges in rotation period: from 1 to 2 days (model \texttt{f0\_1\_2}), from 2 to 4 days (model \texttt{f0\_2\_4}), from 4 to 8 days (model \texttt{f0\_4\_8}), from 8 to 16 days (model \texttt{f0\_8\_16}), and from 16 to 32 days (model \texttt{f0\_16\_32}). In the end, we chose the model with the highest log likelihood and used the posterior PDFs from that model. This is indicated as a ``best model" column in the final catalog and in Table \ref{maintable}. In cases where the posterior PDF of the preferred model peaks at or near a prior boundary and the difference in log probability between the two highest-likelihood models is less than 4, the analysis is repeated with a prior centered at the midpoint between the relevant boundaries, ensuring that the adopted period is not artificially biased by the prior edges.

\subsection{The gyrochrone fitting model}
\label{sec:mixture}
We developed a Bayesian mixture model to describe the relationship between stellar rotation periods and effective temperatures.

We modeled the observed rotation periods as arising from a mixture of two components: a primary gyrochronological relationship and a background population of stars that do not follow the standard age-rotation-temperature relation.

Our gyrochronology model employs Legendre polynomials as the basis functions for regression analysis. While these polynomials are orthogonal on the domain $-1 \leq x \leq 1$, they provide superior numerical stability compared to standard polynomial bases even on unbounded domains. The Legendre polynomials are constructed using the recursion relation:

\begin{equation}
(\ell + 1) P_{\ell+1}(x) = (2\ell + 1) x P_{\ell}(x) - \ell P_{\ell - 1}(x)
\end{equation}

We model the gyrochronological relationship as a function that provides the expected rotation period at any effective temperature:

\begin{equation}
\mu_P(T_\mathrm{eff})
\end{equation}

The true rotation periods follow a normal distribution around this expected relationship with intrinsic scatter $\sigma$:
\begin{equation}
P \sim \mathcal{N}(\mu_P(T_{\mathrm{eff}}), \sigma)
\end{equation}

To ensure numerical stability, we standardize both observed periods $P_\mathrm{obs}$ and temperatures $T_\mathrm{obs}$ to zero mean and unit standard deviation before fitting, then transform results back to physical scales. In this standardized coordinate system, the expected period-temperature relationship is expressed as:
\begin{equation}
\mu_y(x) = \sum_{\ell} C_\ell P_\ell(x)
\end{equation}
where $C_\ell$ are the polynomial coefficients to be inferred.

To account for period measurement uncertainties, we introduce the measurement
model relating the true periods $P$ to the observed periods, 
\begin{equation}
P_\mathrm{obs} \sim \mathcal{N}(P, \sigma_\mathrm{obs}).
\end{equation}

Since this model is linear in $P$, we can analytically marginalize over the latent period variables (see \citet{Hogg2020}):

\begin{equation}
P_\mathrm{obs} \sim \mathcal{N}(\mu_P(T_\mathrm{eff}), \sqrt{\sigma_\mathrm{obs}^2 + \sigma^2})
\end{equation}

To accommodate stars that deviate from the standard gyrochronological sequence, we implement a mixture model where each star's true period probabilistically follows either the gyrochronological relation or a background distribution. With probability $f_\mathrm{bg}$, a star follows the background model:

\begin{equation}
P \sim \mathcal{N}(\mu_\mathrm{bg}, \sigma_\mathrm{bg})
\end{equation}

With probability $(1-f_\mathrm{bg})$, it follows the gyrochronological relation:

\begin{equation}
P \sim \mathcal{N}(\mu_P(T_\mathrm{eff}), \sigma)
\end{equation}

Then the mixture model probability density is:

\begin{equation}
p(P \mid \ldots) \propto (1 - f_\mathrm{bg}) \mathcal{N}[\mu_P(T_\mathrm{eff}), \sigma](P) + f_\mathrm{bg} \mathcal{N}[\mu_\mathrm{bg}, \sigma_\mathrm{bg}](P)
\end{equation}

We analytically marginalize over $P$ in each mixture component and constrain $0 \leq f_\mathrm{bg} \leq 1/2$ to ensure the gyrochronological component remains dominant.

 We also account for uncertainty in temperature measurements that were taken from the \textit{Gaia DR3} catalog \citep{gaiadr3}; the gyrochronological relation above is expressed in terms of the true effective temperature, to which we apply a Normal prior, 
\begin{equation}
T \sim \mathcal{N}(\mu_T, \sigma_T).
\end{equation}
We assume that the observed temperatures are normally distributed around true
temperatures:
\begin{equation}
T_\mathrm{obs} \sim \mathcal{N}(T, \sigma_{T,\mathrm{obs}})
\end{equation}
with $\sigma_T$ as reported in \textit{Gaia DR3} \citep{gaiadr3}. The gyrochronological relation depends on the true temperature
$T$, creating nonlinear dependencies that require direct sampling over the
latent temperature variables rather than analytical marginalization.

Using the ratio of terms in the mixture model, we can compute for each star the
posterior probability that it belongs to the gryrochronoligcal sequence as
opposed to the background population.  Estimates of stars' true $P$ and
$T_\mathrm{eff}$ are improved by the model compared to the raw measurement
error, particularly for stars with high probability of belonging to the
gyrochronological sequence in regions where the sequence is a strong function of
temperature (where our accurate inferences of the rotation period allow to infer
the temperature from the sequence); see Figure \ref{fig:finalgyro}.

\subsection{Vetting} \label{sec:vetting}

\begin{figure*}[ht]
    \center
    \includegraphics[width=\textwidth]{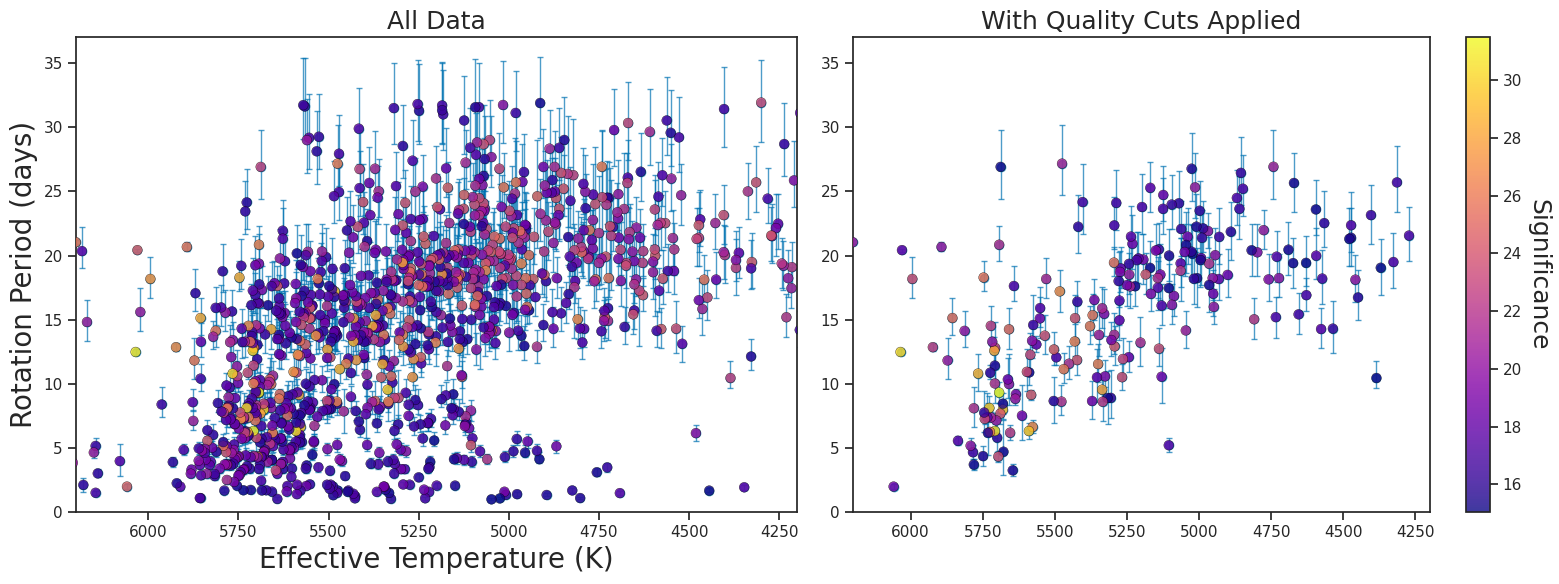}
        \caption{Two plots showing before and after applying the quality cuts described in Section \ref{sec:vetting}. The left panel shows the complete dataset of detected rotation periods, while the right panel displays the same data after applying quality cuts. Quality cuts include: (1) a significance threshold requiring $\text{mean}_\sigma / \text{std}_\sigma$ to ensure sufficient signal-to-noise in the light curve amplitude, (2) a precision cut excluding targets with period uncertainties exceeding 3 days, and (3) a color-magnitude cut to remove evolved giant stars. Data points are colored by the statistical ``significance" as we define in Section \ref{sec:vetting}.}
        \label{fig:qualitycut}
\end{figure*}

From our membership list and complete sample of detected periods, we applied a series of quality cuts to ensure robust rotation period measurements. First, we implemented a ``significance" cut based on the parameter $\sigma$, which represents the RMS fluctuation in the light curve attributable to the rotation terms in the GP kernel and is therefore directly related to the variability amplitude (see Section \ref{sec:gp}). We define the ``significance" as the ratio $\text{mean}_\sigma / \text{std}_\sigma$, and retained only objects with significance $\geq 12$ to ensure sufficient signal-to-noise in the rotational modulation. This value was chosen based on visual inspection of the stellar population in the first panel of Figure \ref{fig:qualitycut}. Second, we imposed a precision requirement by excluding all targets with period uncertainties exceeding 3 days, maintaining a sample with well-constrained rotation periods suitable for our analysis. Figure \ref{fig:qualitycut} illustrates the effectiveness of our quality cut procedure in refining the rotation period sample, comparing the complete dataset of detected periods (left panel) with the filtered sample after applying significance, precision, and giants cuts (right panel).

\begin{figure}[hbt!]
    \center
    \includegraphics[width=0.46\textwidth]{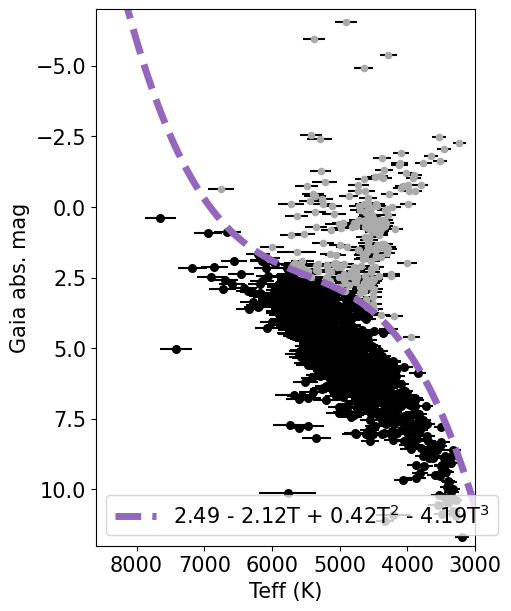}
        \caption{Color-magnitude diagram of 2,622 NGC~6819 members with available values for $T_{\text{eff}}$, Gaia parallax, and Gaia magnitude. The dashed line represents a polynomial fitted to all points brighter than 0th absolute magnitude, with 1.5 magnitudes subtracted to align it with the visually identified upper envelope of the main sequence. The black points are main sequence members identified as acceptable candidates for stellar surface rotation detection, and the 238 grey points were removed from the sample.}
        \label{fig:giantscut}
\end{figure}

To remove evolved stars from our sample, we employed a color-magnitude cut based on the Gaia color-magnitude diagram. We first calculated absolute G magnitudes using the Gaia G magnitudes and parallax measurements \citep{gaiadr3}. For all stars with absolute G magnitude $> 0$, we fit a third-order polynomial of the form:
\begin{align}
\label{eq:giants}
M_G &= 2.49 + (-2.12) \cdot (G_{BP} - G_{RP})  \nonumber \\
&\quad + 0.42 \cdot (G_{BP} - G_{RP})^2 + (-4.19) \cdot (G_{BP} - G_{RP})^3
\end{align}
To define the main sequence upper envelope, we subtracted 1.5 magnitudes from this polynomial fit based on visual identification of a boundary that excludes evolved stars while retaining main-sequence objects. Stars falling above this adjusted boundary were removed from our sample. This cut was applied only to stars with available Gaia G magnitude, parallax, and effective temperature ($T_{\text{eff}}$) measurements. Figure \ref{fig:giantscut} demonstrates our evolutionary status selection using the Gaia color-magnitude diagram to distinguish between main-sequence stars (retained in the sample) and evolved giants (excluded from the sample).

Finally, we sought to ensure that each measured rotation period is correctly associated with its target by ruling out signal contamination. Contamination is a common issue in \textit{Kepler} data owing to the 4'' square pixel size, and presents an increased risk in NGC~6819 due to the high visual density of an open cluster. A contaminant either introduces a spurious variable signal or non-variable photon noise that dilutes the signal from a variable target. We focus on identifying spurious signals, as image subtraction photometry removes non-variable signal as background. Depending on the chosen aperture, a target may also exhibit contamination from bright targets more than a pixel away. We assume that there is minimal contamination from field stars, and, given that many of the targets in this cluster and its field are so faint, we can also assume there is no contamination from stars not in the \textit{Kepler} Input Catalog (KIC). We based our contamination detection on visual inspection of co-trended IRIS light curves for 2,217 cluster members. We performed a cross-inspection of each target's light curve with all light curves found within a radius of 40'', or approximately ten pixels on either side. This is sufficient to capture contamination even from a bright, saturated source at a decently large visual separation from the target. Where multiple targets seemed contaminated by the same signal, we also inspected Lomb-Scargle periodograms \citep{astropy2013}. Each detection of rotation was visually cross-checked with all neighboring light curves in the dataset. We localized each signal to the light curve where it is exhibited with the lease scatter, i.e., where our aperture has captured the maximal photon yield from this target to best resolve the signal, which is possible due to the image subtraction and aperture selection methodology used to produce this uniform dataset. This places reasonable constraints on contamination to within a 4''-square Kepler pixel.
% Image subtraction photometry removes contaminating photon noise as background signal, so we are able to say that typically the highest-amplitude manifestation of any one signal is the real one.

After quality and giant cuts, we had 370 targets for visual inspection. There were only three targets where the detected period was clearly not due to stellar rotation: an eclipsing binary signal in the light curve of KIC~5023951, and the known $\gamma$-Doradus variable KIC~5024084, which also contaminated the light curve of KIC~5024091. This implies that the quality cuts were largely effective in removing detections of binarity or other stellar variability from the sample. Additionally, we removed all stars with RUWE $>$ 1.4 from the \textit{Gaia DR3} catalog. \sabinasays{RUWE is sensitive only to companions at roughly 2–20 AU at the distance of NGC 6819, so closer and wider unresolved binaries remain in the sample}. Of the remaining 367 targets, we identified 45 clearly contaminated light curves and the relevant source of contamination; 38 ``low visibility'' targets where the rotational variability could not be clearly identified by eye; four targets where low data quality (due to systematic detector edge effects that manifest in dramatically different pixel sensitivities between quarters) suggested an inaccurate detection; and six targets affected by both low visibility and data quality issues. We note that the low-visibility detections could be due to contamination by field stars, but regardless of their cause, they are omitted from the remainder of the study. Of the 319 clear signals with no data quality issues, this works out to a contamination fraction of 14\%, which is in line with expectations for being higher than the field contamination fraction of about 11\% \citep{Colman2017}. Our final sample contains 152 reliable detections.

\section{Results \& Discussion} \label{sec:results}

\begin{figure*}[ht]
    \center
    \includegraphics[width=\textwidth]{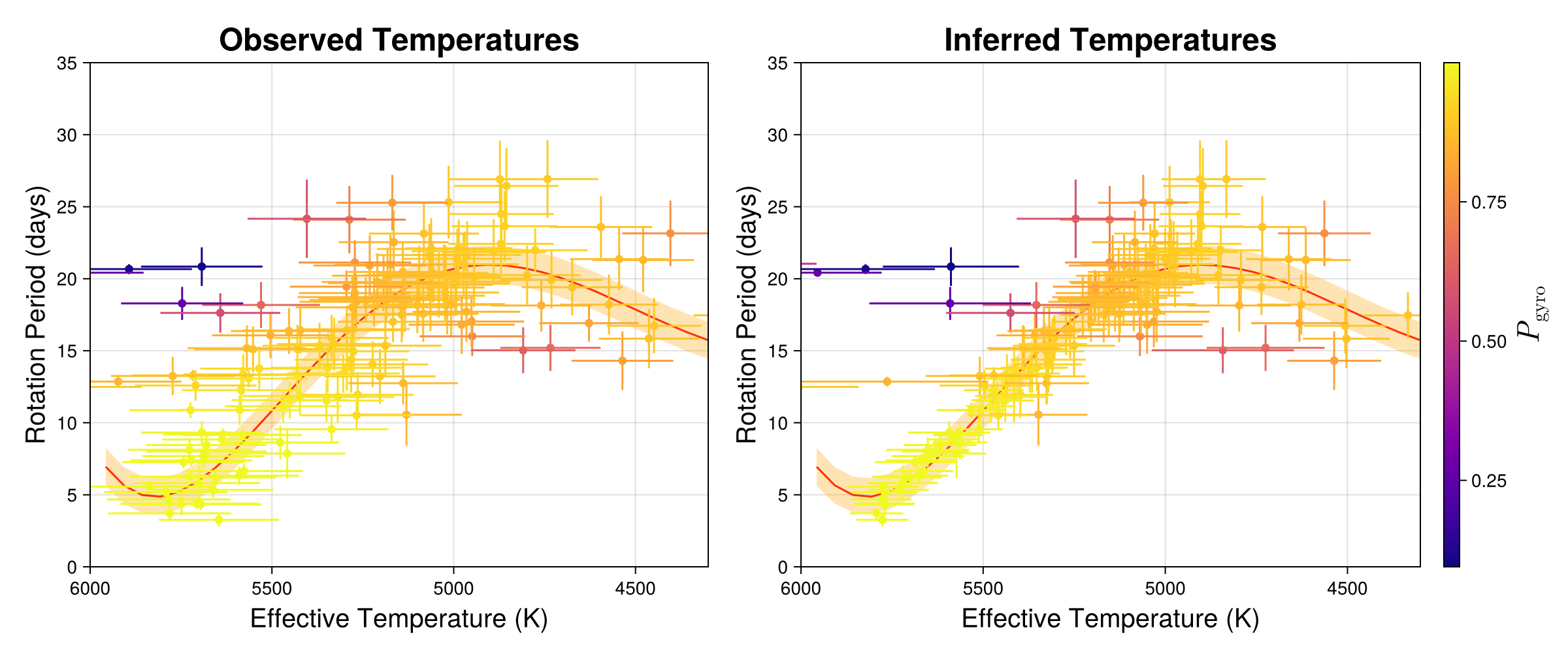}
        \caption{Gyrochronological relationship between stellar rotation period
        and effective temperature. The left panel shows the model fit using
        observed temperatures, while the right panel illustrates inferences of
        the true temperatures as described in Section \ref{sec:mixture}. In all
        cases, errobars give the 1-$\sigma$ (68\%) credible interval.  The red
        curves represent the average of 100 samples with $1\sigma$ (intrinsic
        scatter) uncertainty bands (yellow shading). Data points are colored by
        their posterior probability of belonging to the gyrochrone ($P_{\rm
        gyro}$), from blue (low probability outliers) to yellow (high
        probability). The mixture model successfully identifies and
        probabilistically weights stars that follow the canonical
        gyrochronological relation versus those in the background population.}
        \label{fig:finalgyro}
\end{figure*}

Figure~\ref{fig:finalgyro} presents our final sample of rotation periods after
applying all quality cuts, plotted against effective temperature in the
gyrochronal color-period diagram. Using the model described in Section
\ref{sec:mixture}, we show the fits using observed temperatures (left panel) and
also the inferred true temperatures (right panel). Both panels show rotation
period as a function of effective temperature, with the red curve representing
the median gyrochronological relationship and the yellow shaded regions
indicating the intrinsic scatter (the $\sigma$ parameter). Data points are
color-coded according to their posterior probability of belonging to the
gyrochronological sequence ($P_{\rm gyro}$), with purple points indicating high
probability of following the gyrochrone and green points representing likely
background objects. Our mixture model successfully identifies stars that deviate
from the canonical gyrochronological relation, probabilistically assigning them
to either the gyrochrone or background population. Stars with high $P_{\rm
gyro}$ values (yellow) cluster tightly along the inferred relationship, while
those with low probabilities (blue) are scattered away from the main trend,
demonstrating the model's ability to distinguish between genuine
gyrochronological sequence members.

Table \ref{maintable} provides a representative sample of our final rotation period catalog, showcasing the precision of our measurements.

\begin{table*}[ht]
\centering
\renewcommand{\arraystretch}{1.3} 
\large 
\caption{Representative entries from our rotation period catalog showing KIC identifiers, best-fitting GP model variants, derived rotation periods with uncertainties, and stellar parameters. The ``Best Model'' column indicates the optimal GP kernel configuration selected for each star based on model comparison criteria. Period uncertainties are reported as median values with $-1\sigma$ and $+1\sigma$ bounds from the posterior distributions. Stellar parameters include effective temperature, surface gravity ($\log g$), and stellar radius in solar units taken from the KIC catalog.}
\label{maintable}
\begin{tabular}{ccccc}
\hline
KIC ID & Best Model & $P_{\rm rot}$ (d) & $T_{\rm eff}$ (K) & RUWE \\
\hline
5113396 & $f0\_16\_32$ & $16.47^{+0.43}_{-0.41}$ & $3707 \pm 111$ & 1.058 \\
5024504 & $f0\_8\_16$  & $13.97^{+2.02}_{-1.35}$ & $4027 \pm 119$ & 1.240 \\
5112436 & $f0\_16\_32$ & $17.46^{+1.62}_{-1.56}$ & $4216 \pm 124$ & 0.976 \\
5200222 & $f0\_16\_32$ & $23.15^{+2.32}_{-2.24}$ & $4404 \pm 132$ & 0.331 \\
5200237 & $f0\_16\_32$ & $16.73^{+2.13}_{-1.72}$ & $4449 \pm 130$ & 1.640 \\
5112366 & $f0\_8\_16$  & $15.84^{+2.07}_{-2.03}$ & $4463 \pm 138$ & 0.328 \\
\hline
\end{tabular}
\end{table*}

Our catalog contributes to the existing catalogs with 152 stars with high-quality rotation period measurements. The combination of our expanded NGC~6819 catalog with the established Hyades sequence of 232 stars from \cite{Douglas2019} builds on the foundations of gyrochronology, providing the community with large-sample gyrochrones for a younger and an older open cluster, at the representative ages of $\sim$600-800~Myr (Hyades) and $\sim$2.5~Gyr (NGC~6819). 

\begin{figure}[ht!]
    \center
    \includegraphics[width=0.46\textwidth]{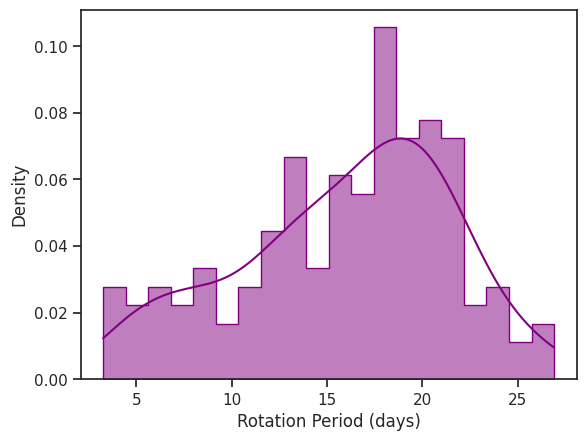}
        \caption{Distribution of rotation periods in our final quality sample. The histogram shows the distribution of measured rotation periods, with the overlaid curve representing a kernel density estimate. The distribution exhibits a peak around 20 days.}
        \label{fig:periodhist}
\end{figure}

Figure~\ref{fig:periodhist} shows the distribution of rotation periods for stars in our NGC~6819 sample. Traditional stellar spin-down theory \citep[e.g.][]{Skumanich1972,Barnes2003} predicts a mass-dependent evolution where more massive, hotter stars (F and early G dwarfs) spin down more gradually due to their thinner convective envelopes and weaker magnetic dynamos, while lower-mass K dwarfs should experience rapid spin-down due to their deeper convective zones and more efficient magnetic braking. Our observed distribution exhibits a long tail with a primary peak around 15-20 days and a significant population of faster rotators with periods between 1-10 days. Because our mixture model assigns stars that deviate from the gyrochronological relation to a background component, unresolved binaries, which tend to appear as fast-sequence outliers, are preferentially excluded from the fitted sequence. This likely mitigates their effect on the slow-sequence scatter.
% We will further discuss the substructure within the primary peak in Section~\ref{sec:pileup}.

\begin{figure*}[ht!]
    \center
    \includegraphics[width=0.9\textwidth]{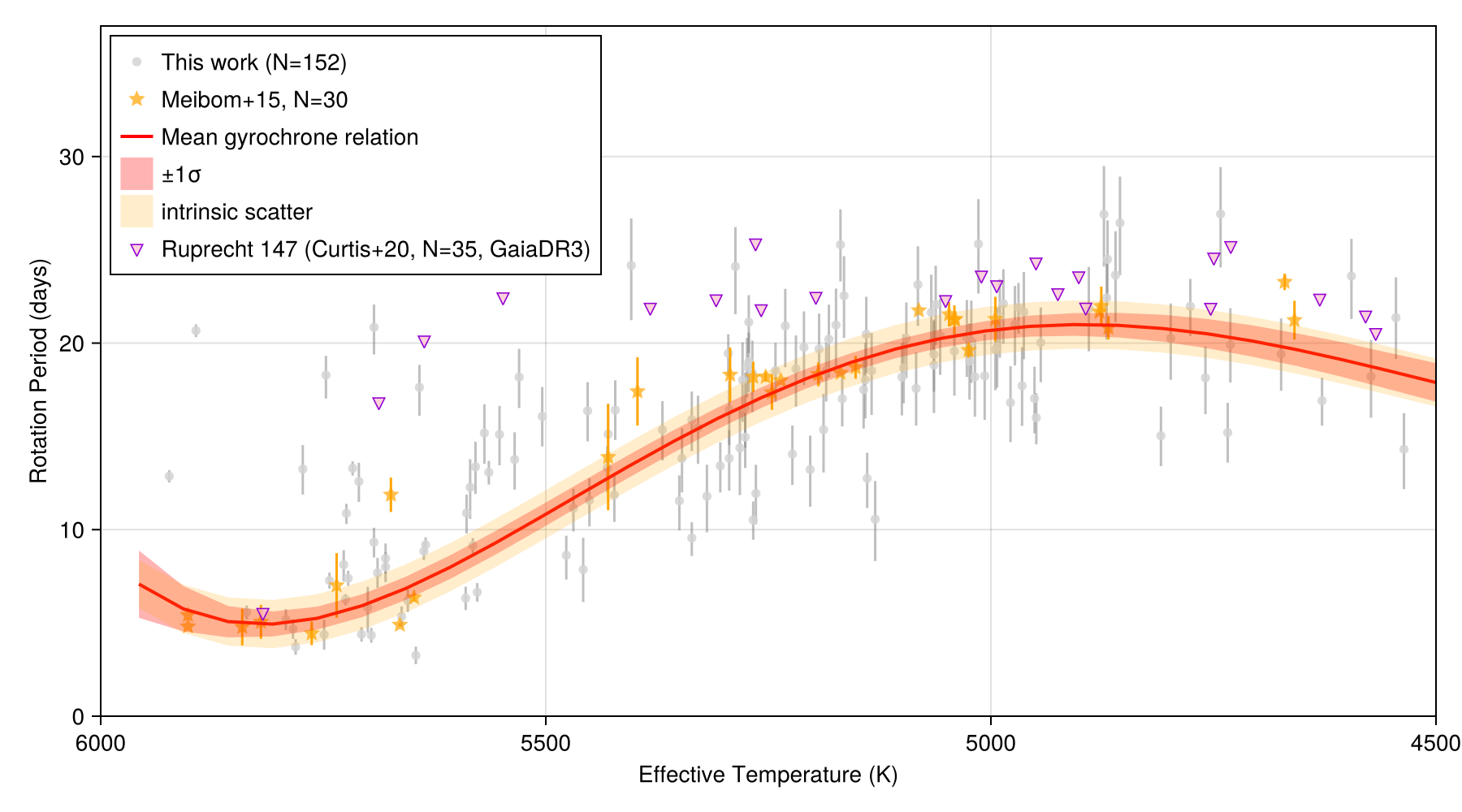}
        \caption{Comparison of our derived gyrochronological relationship with literature estimates. Our \textit{Kepler} field sample (gray circles, N=152) is shown alongside rotation periods from open clusters:  Ruprecht 147 (purple triangles, \cite{Curtis2020}, N=35), and \cite{Meibom2015} measurements (orange stars, N=30). The red curve represents our inferred mean gyrochronological relation of 100 samples, with intrinsic scatter indicated by the yellow shaded region. The agreement between our field star calibration and cluster benchmarks validates our approach.}
        \label{fig:gyrofit}
\end{figure*}

Figure~\ref{fig:gyrofit} shows the rotation period-temperature relation derived from our Bayesian analysis. The orange solid line represents the mean gyrochronological relation obtained from the Legendre polynomial fit, with the orange shaded region indicating the $\pm 1\sigma$ of the curve. Our sample (gray circles) exhibits the expected trend of increasing rotation periods toward cooler temperatures.

We compare our results with previous literature on NGC~6819: the measurements from \cite{Meibom2015} (orange stars with error bars), and cluster members that were represented in the \cite{Santos2021} survey of all Kepler data (red diamonds). We also include the \cite{Curtis2020} results for Ruprecht 147 (purple triangles), which provides a comparison at a similar age of $\sim$2.7 Gyr. As expected, the Ruprecht~147 data points lie towards the upper envelope of our sample.

\begin{figure}[ht!]
    \center
    \includegraphics[width=0.46\textwidth]{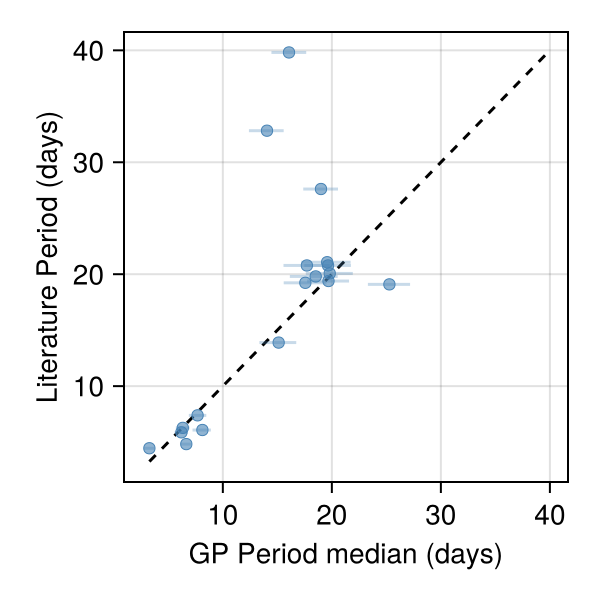}
        \caption{Comparison of our sample of inferred rotation periods to the ones reported in literature by \cite{Meibom2015} and \cite{Santos2021}. The x-axis shows the inferred period from the model in this work, and the y-axis shows the literature periods. The error bars represent the 1-sigma errors from our model. The points that lie on the dashed line agree with the literature values. The values that do not agree are the ones from \cite{Santos2021} catalog, and the light curves for them are shown in Appendix \ref{sec:apxlc}.}
        \label{fig:litfit}
\end{figure}

The literature periods generally follow our derived mean relation, particularly in the temperature range 5500-6000 K, where the gyrochronological sequence is well-established. More comparison to the literature values is shown in Figure \ref{fig:litfit}.

\subsection{Gyrochronology reliability}
Our in-depth study of NGC~6819 presents an opportunity to probe the observational challenges to traditional gyrochronology. This builds on the work by \cite{Curtis2020} with the closely co-eval clusters NGC~6819 and Ruprecht~147. The breakdown of gyrochronology is particularly evident in the apparent ``stalling" of rotational evolution observed in intermediate-age clusters, observed by \cite{Curtis2020} in their combined $\sim$2.5~Gyr dataset. \cite{Douglas2019} found that low-mass stars ($\lesssim 0.6$ M$_{\odot}$) in the 700 Myr Praesepe cluster rotate at essentially the same rate as their counterparts in the $\sim$670 Myr Praesepe cluster, despite the age difference. Separately, \cite{Curtis2019} found similar convergence between the 1 Gyr NGC 6811 cluster and younger Hyades/Praesepe. 

To further contextualize our observations, we turn to the work by \cite{Spada2020}, who demonstrated that the competing effects of surface magnetic braking and interior angular momentum transport can naturally explain the observed stalling phenomenon. The efficiency of the internal angular momentum transport process exhibits a steep mass dependence, causing different mass stars to undergo re-coupling between their radiative cores and convective envelopes at different ages. During the re-coupling phase, the apparent surface spin-down is dramatically reduced or ``stalled" because the angular momentum transported from the interior partially compensates for the angular momentum loss at the surface due to the magnetized wind. The \cite{Spada2020}
model predicts large ranges in rotation period across stellar mass for young coeval groups, but it is precisely around 2--3~Gyr that the model starts to converge for masses in the FGK regime. To return to the example from \cite{Curtis2019}, this effect naturally explains why stars in the young clusters NGC 6811 and Praesepe show overlapping rotation periods at low masses while maintaining the expected age ordering at higher masses (where re-coupling occurred earlier).

The challenges are not limited to individual stellar evolution but extend to population-level analyses. \cite{Lu2021} examined rotation period distributions in different Galactic populations and found that the relationship between rotation and kinematics -- a proxy for age -- varies systematically with stellar mass and metallicity in ways that simple gyrochronology models cannot reproduce. This suggests that environmental factors, stellar composition, and possibly even Galactic chemical evolution history may influence rotational evolution in ways not currently accounted for in age-dating prescriptions.
Additionally, it is important to note that all theoretical gyrochronology calibrations, including those discussed here, assume single-star evolution, but a significant fraction of observed field stars may have experienced binary interactions that would alter their rotation rates in ways not captured by single-star models. We have not removed photometric or spectroscopic binaries. Binaries are over-represented among rapid outliers in open clusters \citep[e.g.,][]{Curtis2020}, so some stars our mixture model assigns to the background population are likely binaries, and the intrinsic scatter we infer should be regarded as an upper limit on that of single stars.

The consensus emerging from recent works is that while rotation periods remain valuable age indicators for young ($<1$ Gyr) stellar populations, their reliability diminishes significantly for older stars, particularly those with rotation periods exceeding $\sim15-20$ days or in mass ranges experiencing rotational coupling transitions.
Accurate gyrochronological age determination requires detailed modeling of both magnetic braking and internal angular momentum transport, with careful consideration of the mass-dependent timescales governing core-envelope coupling and the potentially variable initial conditions set during the pre-main-sequence phase. 

Our results for NGC 6819 ($\sim$2.5 Gyr) provide direct observational evidence for these theoretical predictions at older ages. The scatter observed in our rotation period measurements, with many stars rotating significantly faster than expected from standard gyrochronological relations, demonstrates that the rotational coupling effects identified by \cite{Spada2020} may continue to influence stellar evolution well beyond 1 Gyr, although a contribution from unresolved binaries cannot be excluded. However, while \cite{Spada2020} demonstrated that rotational coupling effects can explain the stalled spin-down observed between Praesepe and NGC 6811, their model predicts that most stars should achieve quasi-solid-body rotation and enter a stable wind-braking-dominated regime by ages of 2-3 Gyr. The substantial rotational scatter we observe with our expanded sample of NGC 6819 periods therefore suggests that additional physical mechanisms beyond the two-zone coupling model may be operating at these older ages.

\section{Conclusions} \label{sec:conclusions}
In this work, we used \textit{Kepler} IRIS light curves to remeasure stellar rotation periods in the open cluster NGC 6819, producing a comprehensive rotation period catalog for an intermediate-age open cluster. By applying a combination of Gaussian Process modeling, rigorous vetting, and membership constraints, we isolated 152 high-confidence rotation period detections from an initial sample of over two thousand stars. This effort expands the available rotator sample in NGC 6819 by an order of magnitude compared to previous studies.

Our results confirm the overall trend of increasing period toward cooler stars, but also reveal significant intrinsic scatter that cannot be reconciled with simple spin-down models. The presence of both fast and slow rotators at the same effective temperature suggests that internal angular momentum transport and mass-dependent coupling between stellar interiors and envelopes play a key role in shaping rotational evolution beyond $\sim$1 Gyr. 

Together, our results reinforce NGC 6819 as a cornerstone for testing gyrochronology at intermediate ages, but also demonstrate that standard gyrochronological relations break down at the 2-3 Gyr timescale. Rotation periods remain powerful tracers of stellar physics, yet their use as precise chronometers for older field stars requires models that incorporate angular momentum transport and magnetic braking models. Our expanded catalog provides a critical benchmark for calibrating such next-generation gyrochronology models, and will serve as a valuable dataset for comparisons with upcoming missions such as \textit{PLATO}.

\section*{Acknowledgments}
We thank the anonymous referee for insightful comments. S.S. thanks Andy Casey for valuable discussions. The authors also thank Diego Godoy Rivera for valuable comments on the first draft. S.S. acknowledges support from award 644616 from the Simons Foundation.  S.S. thanks Kevin Hardegree-Ullman for cross-matching Gaia DR3 IDs and KIC IDs.  Software citation information aggregated (in part) using \texttt{\href{https://www.tomwagg.com/software-citation-station/}{The Software Citation Station}} \citep{software-citation-station-paper, software-citation-station-zenodo}.


% \vspace{5mm}
% \facilities{HST(STIS), Swift(XRT and UVOT), AAVSO, CTIO:1.3m,


\software{\texttt{Astropy} \citep{astropy2013, astropy2018, astropy2022},
\texttt{Matplotlib} \citep{matplotlib},
\texttt{NumPy} \citep{numpy},
\texttt{Pandas} \citep{pandas},
\texttt{SciPy} \citep{scipy},
\texttt{PyMC} \citep{pymc},
\texttt{emcee} \citep{emcee},
\texttt{julia} \citep{bezanson2017julia},
\texttt{Turing.jl } \citep{ge2018t},
\texttt{Makie.jl} \citep{Makie},
\texttt{ArviZ} \citep{arviz_2019},
\texttt{xarray} \citep{hoyer2017xarray}.
}

\textit{Data availability}: The final catalog and the model implementations are all available on Zenodo\footnote{see \url{https://doi.org/10.5281/zenodo.17247991}}



\appendix
\section{A random sample of light curves}
\label{sec:apxlc}
Here, we present a sample of inferred periods from our catalog. They are represented in three panels. The first panel shows the posterior distribution for periods; the second panel shows the full co-trended light curve; the third panel shows only the 200-day chunk of the full light curve, with blue lines showing the chunks of the light curve separated by the median of the period posterior distribution. We show each quarter's flux relative to \emph{that quarter's} mean flux. Therefore, the apparent changes in the fluctuations quarter-to-quarter are not necessarily real, as the mean flux will vary from quarter to quarter based on the instrument sensitivity.  The model described in Section \ref{sec:gp} accounts for this, assigning the (common-to-all-quarters) GP to the fractional flux variations, with a per-quarter mean flux that is fit to the data. In some of the samples, we also show the rotation period inferred in the literature by \cite{Santos2021}; these are shown in red lines.

The next Figure shows similar panels with an inclusion of posterior distributions for $\sigma$. The goal here is to illustrate two lowest-significance light curves, two median-significance light curves, and two highest-significance light curves. The time interval of the zoomed-in light curves was chosen for no scientific reason. We provide a notebook where one can plot the same figure using our data.

\begin{figure}[ht!]
    \center
    \includegraphics[width=0.8\textwidth]{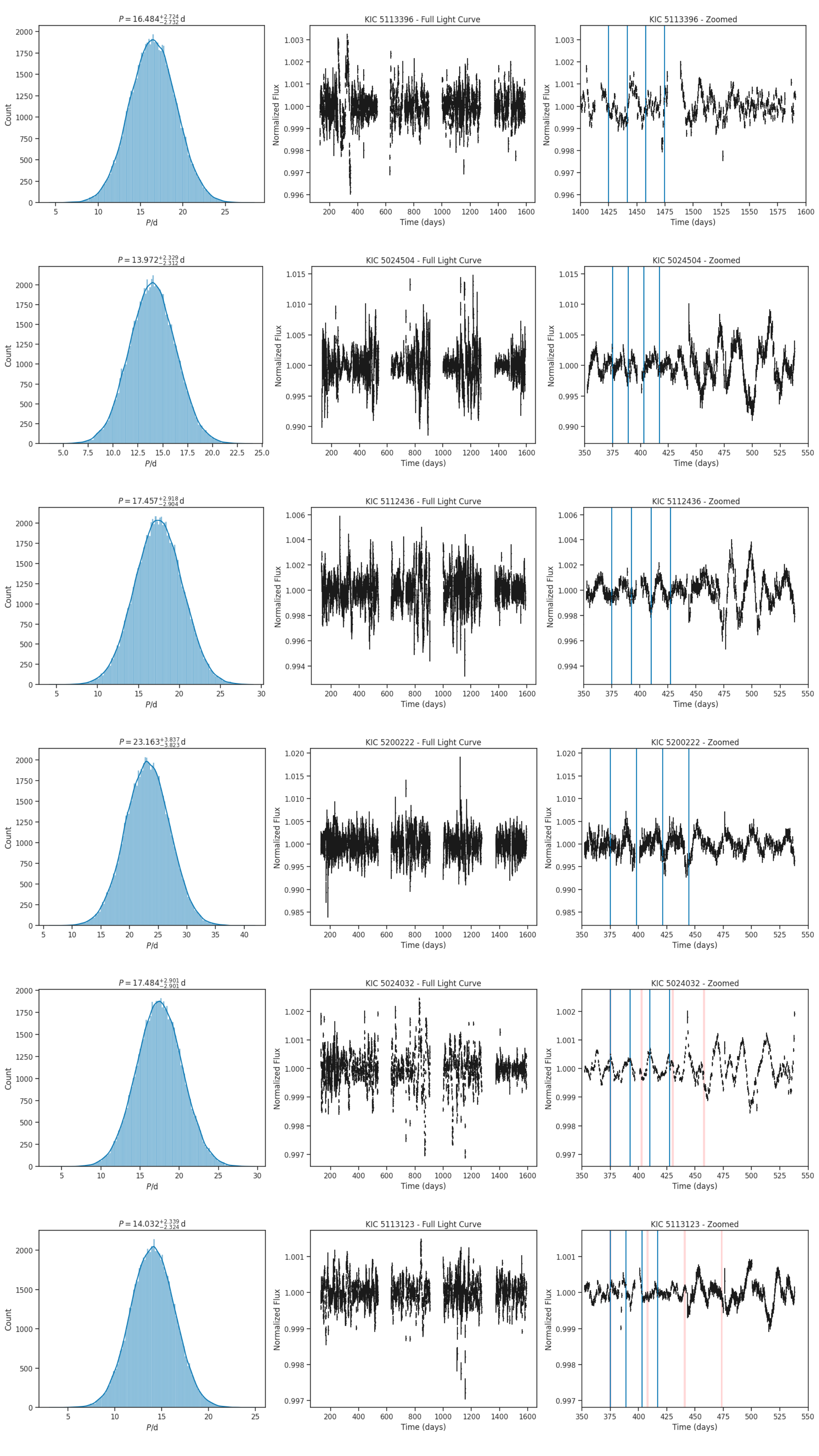}
        % \caption{Distribution of rotation periods in our final quality sample. The histogram shows the distribution of measured rotation periods, with the overlaid curve representing a kernel density estimate. The distribution exhibits a peak around 20 days.}
        \label{fig:periodsamples}
\end{figure}
\begin{figure}[ht!]
    \center
    \includegraphics[width=0.8\textwidth]{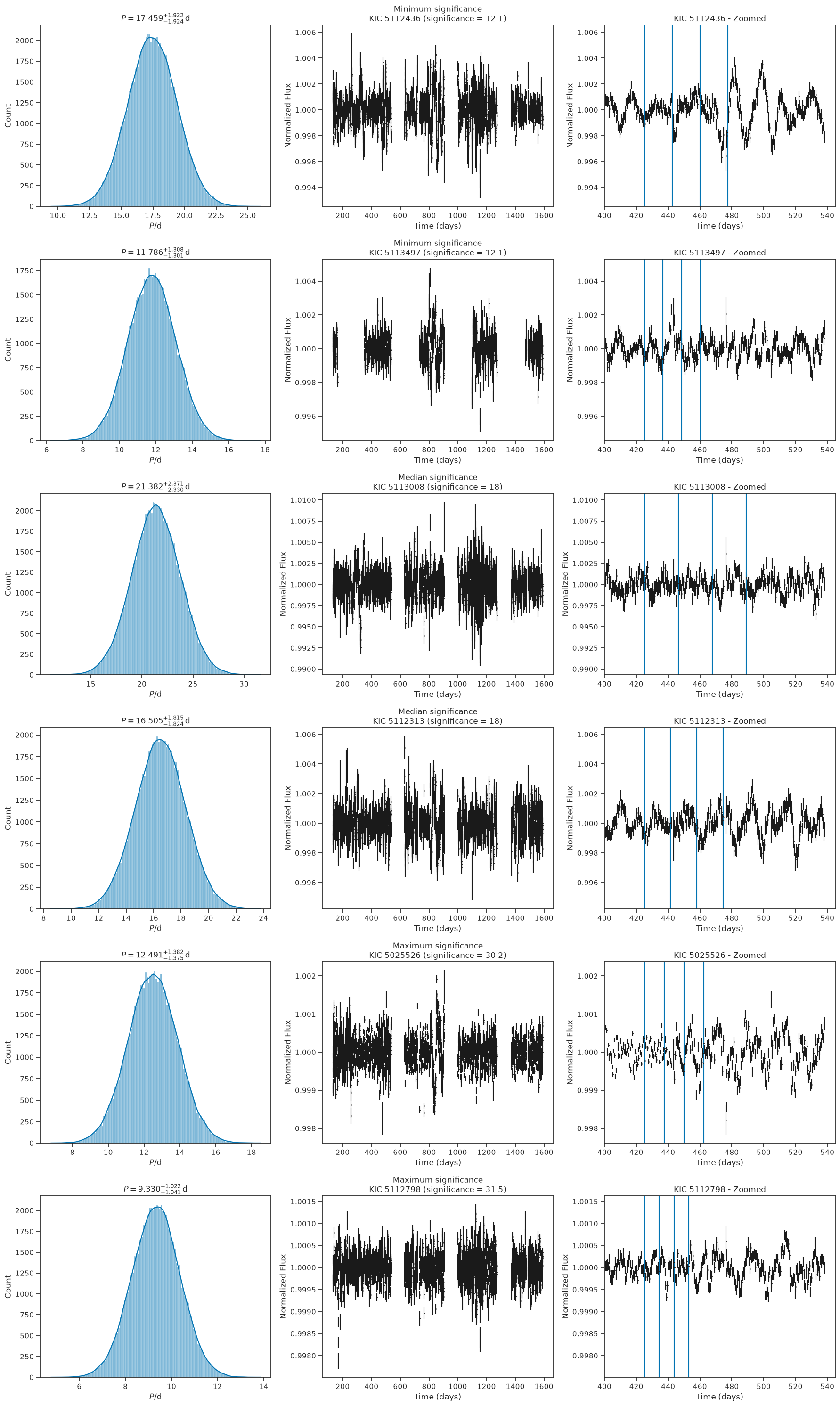}
        % \caption{Distribution of rotation periods in our final quality sample. The histogram shows the distribution of measured rotation periods, with the overlaid curve representing a kernel density estimate. The distribution exhibits a peak around 20 days.}
        \label{fig:periodsamples}
\end{figure}

\newpage
\section{CBVs for each quarter}
\label{sec:cbvs}
\begin{figure}[ht!]
    \center
    \includegraphics[width=0.3\textwidth]{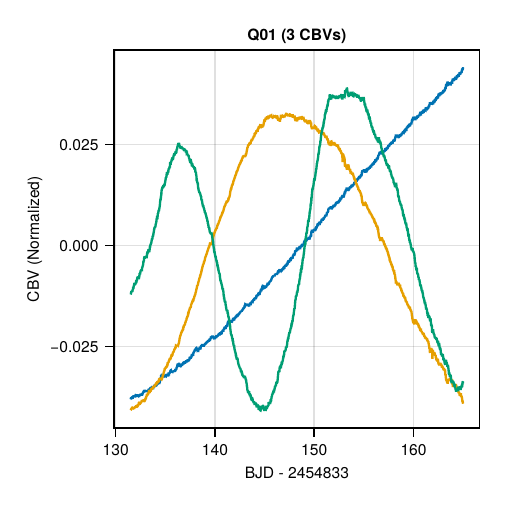}
    \includegraphics[width=0.3\textwidth]{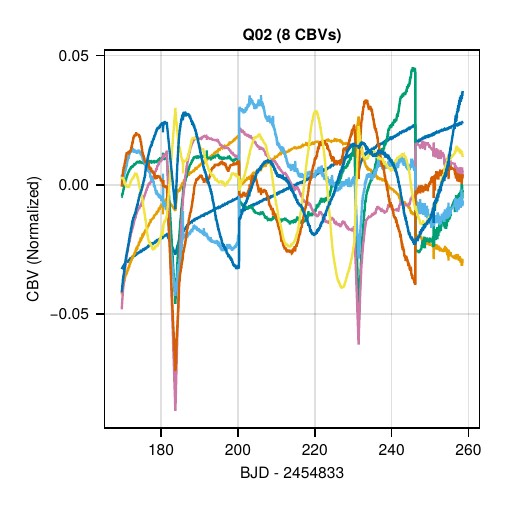}
    \includegraphics[width=0.3\textwidth]{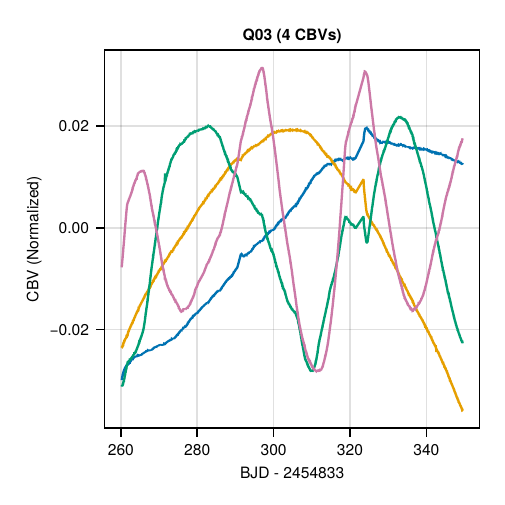}
    \includegraphics[width=0.3\textwidth]{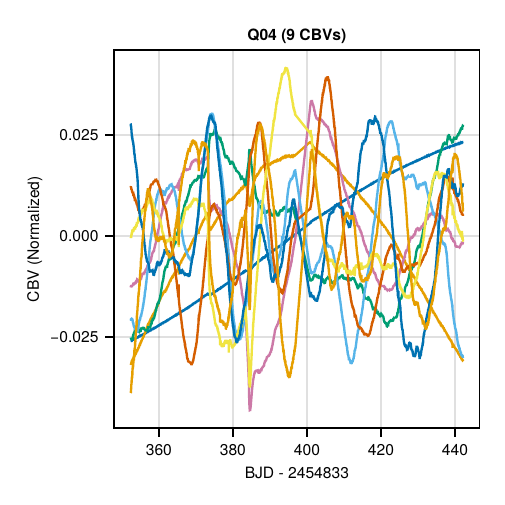}
    \includegraphics[width=0.3\textwidth]{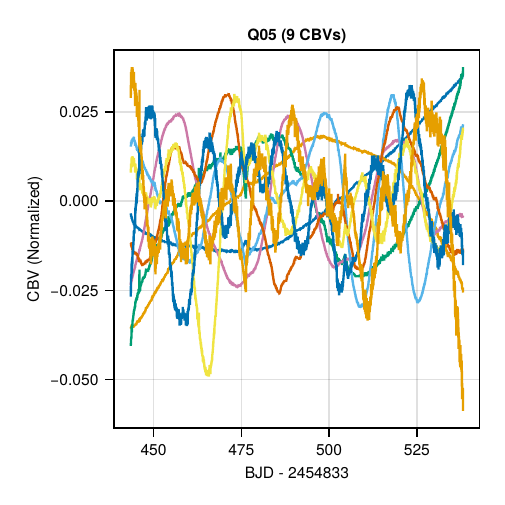}
    \includegraphics[width=0.3\textwidth]{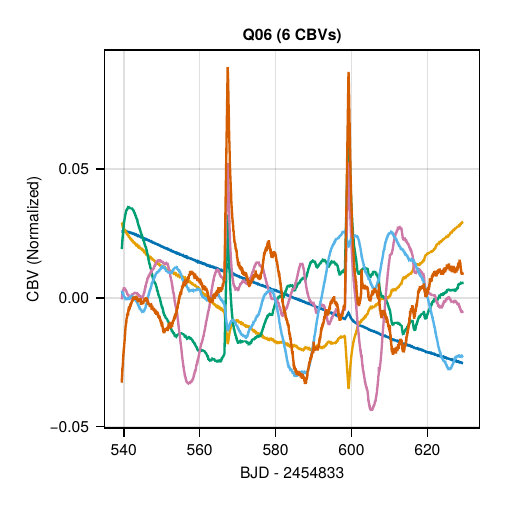}
    \includegraphics[width=0.3\textwidth]{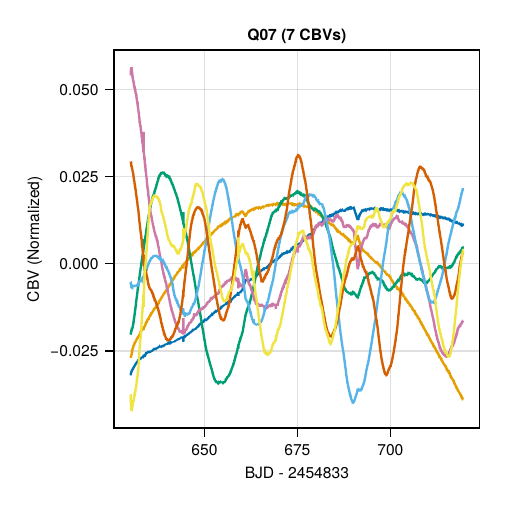}
    \includegraphics[width=0.3\textwidth]{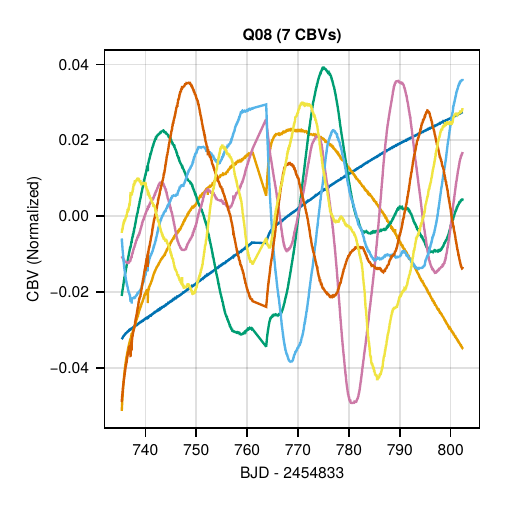}
    \includegraphics[width=0.3\textwidth]{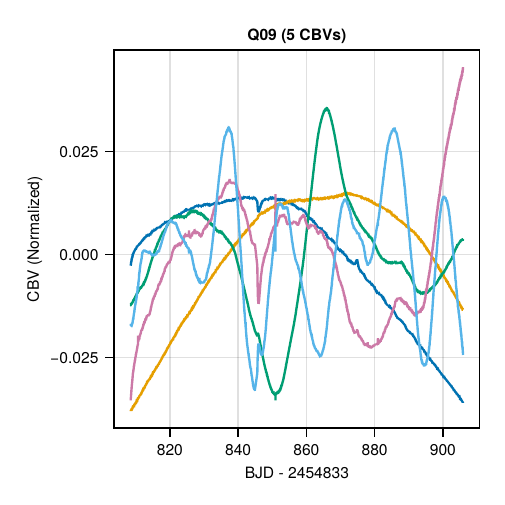}
    \includegraphics[width=0.3\textwidth]{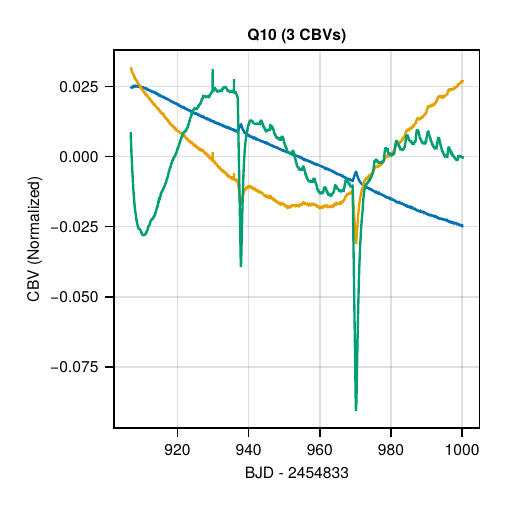}
    \includegraphics[width=0.3\textwidth]{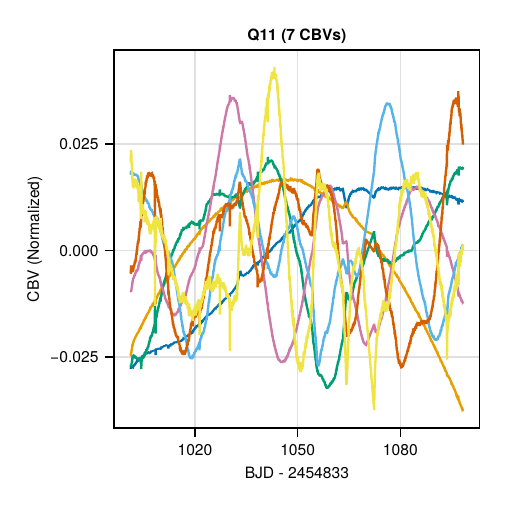}
    \includegraphics[width=0.3\textwidth]{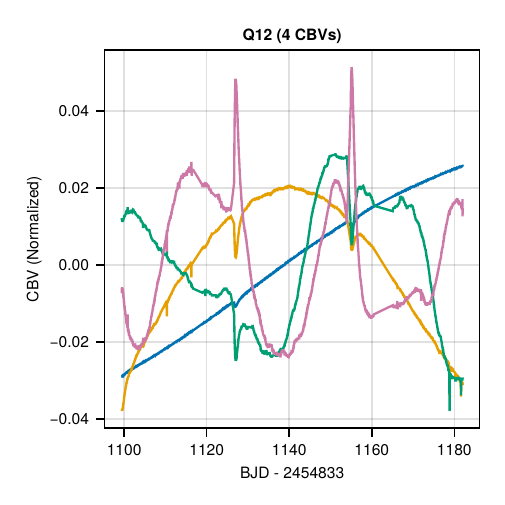}
    % \includegraphics[width=0.3\textwidth]{Q13_CBV.pdf}
    % \includegraphics[width=0.3\textwidth]{Q14_CBV.pdf}
    % \includegraphics[width=0.3\textwidth]{Q15_CBV.pdf}
    % \includegraphics[width=0.3\textwidth]{Q16_CBV.pdf}
    % \includegraphics[width=0.3\textwidth]{Q17_CBV.pdf}
        % \caption{}
        \label{fig:QCBVs1}
\end{figure}
\begin{figure}[ht!]
    \center
    \includegraphics[width=0.3\textwidth]{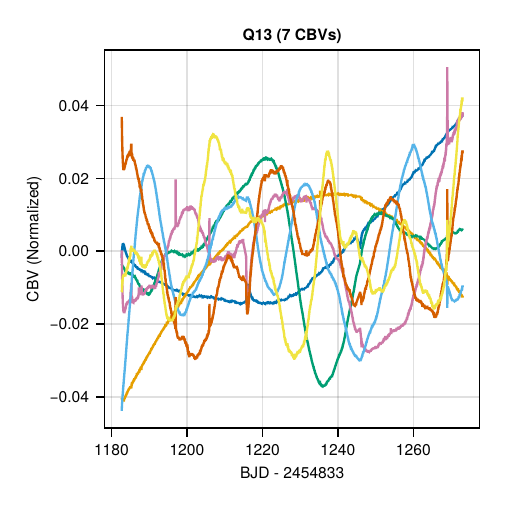}
    \includegraphics[width=0.3\textwidth]{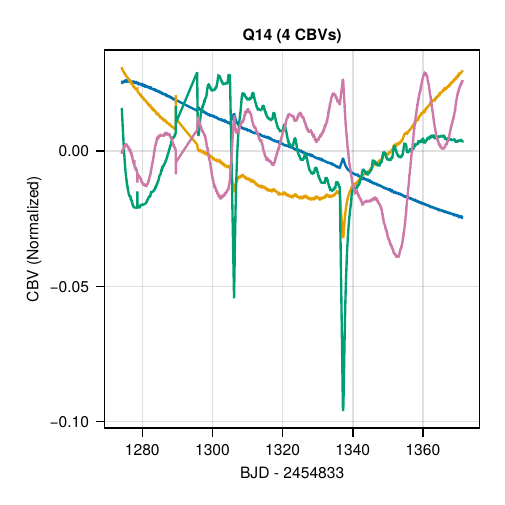}
    \includegraphics[width=0.3\textwidth]{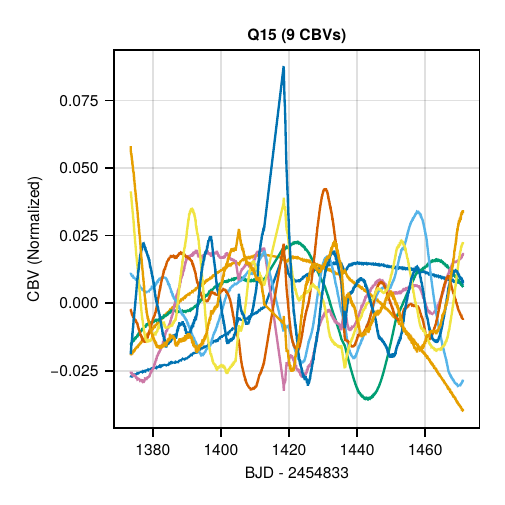}
    \includegraphics[width=0.3\textwidth]{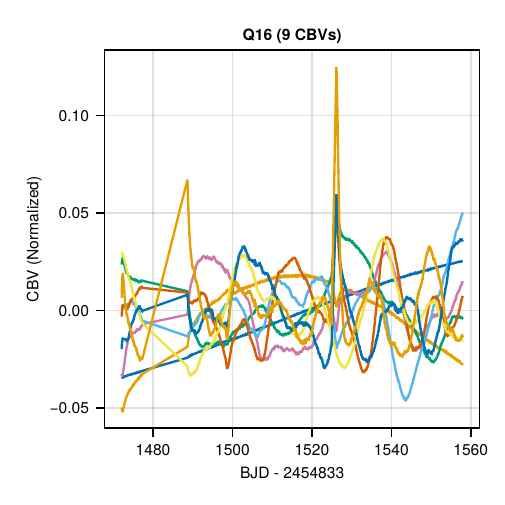}
    \includegraphics[width=0.3\textwidth]{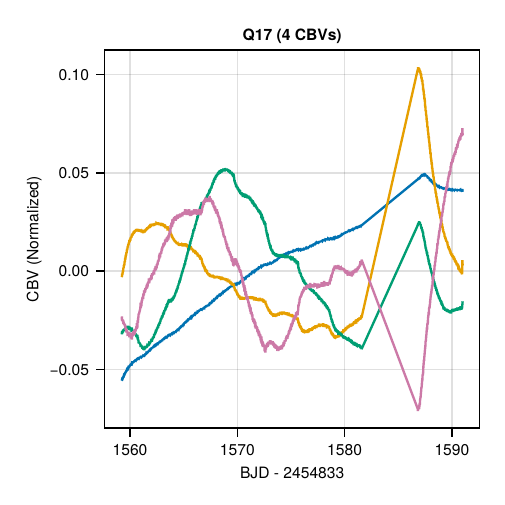}
        \caption{CBVs used for all quarters of the data. Different colors represent different CBVs. The number of CBVs was determined by includeing additional CBVs until the coefficient of the last-added CBV in the median star reached 5-sigma significance above zero. }
        \label{fig:pQCBV2}
\end{figure}

\newpage
\bibliography{references}{}
\bibliographystyle{aasjournal}

\end{document}